\documentclass[a4paper,12pt]{article}
\usepackage{jheppub}
\usepackage{dcolumn}% Align table columns on decimal point
\usepackage{bm}% bold math
\usepackage{esvect}
\usepackage{accents}

\usepackage{hyperref}
\hypersetup{colorlinks=true,linkcolor=magenta,anchorcolor=green,citecolor=cyan,filecolor=black,menucolor=black,urlcolor=brown}
\newcommand{\ba}{\begin{eqnarray}}
\newcommand{\ea}{\end{eqnarray}}
\newcommand{\Mp}{M_{\textrm{Pl}}}
\newcommand{\mpl}{m_{\textrm{Pl}}}
\newcommand{\be}{\begin{equation}}
\newcommand{\ee}{\end{equation}}
 \usepackage[compat=1.1.0]{tikz-feynman}
\usepackage{tikz,contour}
\usetikzlibrary{calc,decorations.markings}
\pgfdeclarelayer{bg}    % declare background layer
\pgfsetlayers{bg,main}  % set the order of the layers (main is the standard layer)

\numberwithin{equation}{section}

\begin{document}

\preprint{IPHT-t23/049, LAPTH-033/23}

\title{Scalaron dynamics from the UV to the IR regime}

\author{Philippe Brax and Pierre Vanhove}
\affiliation{Institut de Physique Th\'eorique, Universit\'e
  Paris-Saclay, CEA, CNRS, F-91191 Gif-sur-Yvette Cedex, France}

\abstract{
We consider a scenario where the scalaron of $f({\cal R})$ models
is related to the  volume modulus of string compactifications  leaving  only one scalar degree of freedom at low
energy. The coefficient of the leading curvature squared
contribution to the low-energy effective action of gravity determines the mass of the scalaron. We impose that this mass is small  enough to allow for  the scalaron to drive Starobinski's inflation.
After inflation, the renormalisation group evolution of the couplings
of the $f({\cal R})$ theory, viewed as a scalar-tensor theory, provides  the link
with the Infra-Red regime.  We consider a scenario where  the
corrections to the mass of the scalaron are large and reduce it below
the electron mass in the Infra-Red, so that the  scalaron plays a central role in the low-energy dynamics of the Universe. In particular this leads to  a connection between  the scalaron mass and the
measured vacuum energy provided its renormalisation group running at energies higher than the electron mass never drops below the present day value of the dark energy.
}

\maketitle

\section{Introduction}
The dark sector of the Universe, comprising both dark matter and dark energy, is as elusive as ever~\cite{Copeland:2006wr,Joyce:2014kja}. One interesting possibility would be that
the dark energy sector could have a complete gravitational origin~\cite{Brax:2017idh,Sotiriou:2008rp}. Usually, this is understood in such a way that gravity, i.e.\ Einstein's General Relativity (GR), should be modified and for instance become massive~\cite{Fierz:1939ix}. This is fraught with conceptual difficulties~\cite{deRham:2014zqa}. On the other hand, a less trodden path could be that gravity is simply modified
by the presence of small corrections to the Einstein-Hilbert action whose role would be to modify the quantum nature of the vacuum. We introduce a scenario whereby the leading Ricci scalar squared  ${\cal R}^2$ corrections dominate at low energy as long as the compactification scale is large enough and the non-perturbative effects involved in the moduli stabilisation occur at low enough energy compared to the compactification scale. This circumvents the problem of hierarchy between the ${\cal R}^2$ and the higher order terms  noticed in~\cite{Brinkmann:2023eph} and give a valid description of the dynamics of the Universe from the inflation scale down to the Infra-Red (IR).  This is the approach we will follow in this paper where we will argue that string compactifications of physical relevance should enrich the gravitational sector of the theory with  such  $f({\cal R})$ corrections~\cite{Brax:2017bcp,Brax:2019iut,Burgess:2016owb}.

Our starting point is the gravitational action of string theory in ten
dimensions. Irrespective of the details of the mechanism, dimensions
must be reduced from ten to four and in this process the universal
volume modulus representing the typical size of the compactification
manifold appears~\cite{Ferrara:1988ff}. After compactification, the low-energy gravitational
action contains terms involving the Riemann tensor of the four
dimensional Universe and the volume modulus. The volume modulus has no
potential and therefore appears as a very dangerous massless field
whose coupling to matter would exceed the bounds from gravitational
physics in the Solar System, e.g.\ the Cassini bound on the Yukawa
coupling of such a light scalar to fermions~\cite{Bertotti:2003rm}. In this paper, we will
not try to generate the non-trivial scalar potentials which could lead
to the screening of the volume modulus in the Solar System~\cite{Brax:2021wcv}. We will
simply assume that the volume modulus is stabilised~\cite{Cicoli:2023opf} and acquires a
given mass which is large enough to evade tests of short range
interactions such as provided by the E\"ot-Wash experiment~\cite{Lee:2020zjt}. Once
stabilised, we find that the gravitational effective action in four
dimensions becomes a $f({\cal R})$ model where the scalaron field~\cite{Starobinsky:1980te,Starobinsky:1981vz}
is then identified with the volume modulus. To do so we restrict ourselves
to ghost-free effective actions and assume that the ghost fields~\cite{Stelle:1977ry} induced by higher derivative operators in the gravitational field have
a mass which is rejected at the cut-off scale of the model, i.e.\ the
compactification scale. This leaves a well-ordered $f({\cal R})$ action
where, in the small curvature regime below the string and
compactification scales, the Ricci scalar squared ${\cal R}^2$  term dominates.

 Written as a scalar-tensor theory depending on the scalaron, and
 imposing the existence of a supersymmetric origin for this low-energy
 action, we find the most general K\"ahler potentials and
 superpotentials which are compatible with the Ricci scalar squared ${\cal R}^2$  structure. They involve a two-parameter family of K\"ahler potentials including the familiar no-scale case and
 a fully determined and unique superpotential.
  Of course, this theory is determined in the Ultra-Violet (UV) at the
 compactification scale. It has all the features to generate
 primordial inflation like in the original Starobinski scenario~\cite{Starobinsky:1980te,Starobinsky:1981vz,Brinkmann:2023eph}. Here
 inflation is determined by the coefficient $c_{\rm UV}$ of the Ricci scalar
 squared ${\cal R}^2$  correction in the UV. From the compactification
 point of view, this is a bottom-up constraint as no preferred value
 of $c_{\rm UV}$ originates from the compactification. Notice that here inflation appears as a consequence of both the volume modulus stabilisation and the fact that it happens in a regime where all the higher-order corrections involving higher powers of the curvature are negligible.

We then use a renormalisation group approach to tie up the regime described above after the end of inflation  and the low-energy regime of the theory. In this Wilsonian approach, the Lagrangian of the theory is determined in particular by the vacuum energy and the coefficient $c$ of the ${\cal R}^2$ term which evolve with the scale as particles are integrated out. In a given energy range $\mu$  corresponding to a given temperature $T$ of the Universe, the Wilsonian action and its coefficients depend on $\mu$.  We take as the initial point of the renormalisation group evolution at the reheat temperature and integrate out particles as $\mu$ is decreased toward low energies.
We analyse the
 renormalisation group evolution of the Ricci scalar squared ${\cal
   R}^2$  theory down to the IR  limit at energies
 well below the masses of all massive particles  present in the Universe. We
 identify this long-distance limit as the vacuum energy which
 engenders dark energy.
 The scalaron mass parameter at a given scale $\mu$  is given by the inverse of the square of the
 coefficient of the ${\cal R}^2$. We assume following~\cite{Brax:2019iut} that the scalaron mass in the IR
 is lower than the electron mass and therefore plays a role in the
 dynamics of the vacuum at low energy. Unfortunately the mass of  the
 scalaron in the IR is not determined completely as it depends on the
 UV properties of the model. The dynamics can only be closed by
 imposing reasonable assumptions on the IR properties of the
 theory. In particular, following the reasoning
 of~\cite{Weinberg:1988cp}, we will require that the cosmological
 constant $\rho_\Lambda (\mu)$ at the energy scale $\mu$ should be
 large enough that any bound structure in the Universe with such an
 energy scale, e.g.\ a gas cloud with a temperature $\mu \simeq T$,
 does not collapse within the age of the Universe. This is guaranteed
 as long as $\rho_\Lambda (\mu)\gtrsim -\rho_\Lambda$ where
 $\rho_\Lambda$ is the dark energy of the Universe. Using these
 constraints for $\mu\simeq m_e$ at the electron mass corresponding to
 the X-ray emitting gas in a galaxy cluster, we found
 in~\cite{Brax:2019iut} that the mass of the scalaron is tightly
 bounded once the E\"ot-Wash experiment bound from gravity tests is taken into account. This almost determines the value of $c_{\rm IR}$ in the IR and therefore using the renormalisation group evolution the cosmological constant at the end of inflation. This set of phenomenological constraints provides a bottom-up approach to the physics from the compactification of string theory down to four dimensions. If the scalaron had very large mass in the IR, i.e.\ larger than the electron mass, then  no such constraints would stand. We also notice that the interval of masses for the scalaron in the IR is compatible with scenarios where the scalaron plays the role of dark matter \cite{Shtanov:2021uif}.

 This paper is arranged as follows. In Sect.~\ref{sec:vol},
 we consider the compactification from ten dimensions to four
 dimensions and the identification of the volume modulus with the
 scalaron. In Sec.~\ref{sec:susy}, we impose that the low-energy
 effective action describing the Ricci scalar squared  ${\cal R}^2$  theory comes from a
 supergravity model and determine the K\"ahler potential and the
 superpotential.  We find that the K\"ahler potential is a two-parameter deformation of the no-scale model describing the volume modulus at tree level. Finally, in Sec.~\ref{sec:quant} we describe the
 renormalisation evolution of the Ricci scalar squared ${\cal R}^2$  model down to the IR.
 We conclude with Sec.~\ref{sec:conclusion}. We give an explicit example of potential for the volume modulus in an appendix.  We also discuss the renormalisation group and the thermodynamic decoupling of scalarons in two appendices.

\section{The volume modulus as scalaron}
\label{sec:vol}

In this section we present a scenario where the Ricci scalar squared
${\cal R}^2$ model leading to Starobinski's inflation is induced from
the compactification of extra dimensions such as the reduction from
ten dimensions  to four dimensions of string theory.
For this we first detail in Sec.~\ref{sec:dimred} how, in the
so-called supergravity frame, the volume modulus does not have a kinetic term.
We then use this in Sec.~\ref{sec:Vscal} for the
scalaron with the universal volume modulus. We discuss volume
stabilisation in Sec.~\ref{sec:Vstab}, and show, in Sec.~\ref{sec:highderivatives} that higher
derivatives corrections induced by string theory are negligible.  We discuss in  \ref{sec:val} the validity of the
quadratic approximation used to derive the Starobinski model during inflation where the excursion of the volume modulus is large~\cite{Cicoli:2015wja}.

\subsection{Dimensional reduction}\label{sec:dimred}

We consider a $4+d$ space-time and compactify on a $d$ manifold of vanishing Ricci scalar~\cite{Otero:2017thw}. We will consider first the case of the Einstein-Hilbert action
\begin{equation}
S_{4+d}= M_{2+d}^{d+2} \int d^{4+d} x \sqrt{-g^{4+d}} R_{4+d}
\end{equation}
where $R_{4+d}$ is the Ricci scalar of the metric $g^{4+d}$. We
consider a  metric of the form
\begin{equation}
g^{4+d}_{ab}=  g^4_{\mu\nu}dx^\mu dx^\nu + \sigma^2(x) g_{ij}^d dx^i dx^j
\end{equation}
 Defining  the volume of the compactification as
\begin{equation}
 { V}_0= \int d^d x \sqrt{g^d},
\end{equation}
 we find after dimensional reduction the effective action
\begin{equation}
S_4= M_{2+d}^{d+2} { V}_0 \int d^4x \sqrt{-g^4}\sigma^{d}  \left(R_4+ d(d-1) (\partial \ln\sigma)^2\right)
\end{equation}
We can go to the Einstein frame  by defining
\begin{equation}
g^4_{\mu\nu}= \sigma^{-d} g^E_{\mu\nu}
\end{equation}
leading to
\begin{equation}
S_4= M_{2+d}^{d+2} { V}_0 \int d^4x \sqrt{-g^E}(  R^E-\frac{d(d+2)}{2} (\partial \ln \sigma)^2).
\end{equation}
This action can be transformed into
\begin{equation}
S_4= M_{2+d}^{d+2} { V}_0 \int d^4x \sqrt{-g} \sigma^{-2\alpha} R
\label{act}
\end{equation}
where
\begin{equation}
g^E_{\mu\nu}= \sigma^{-2\alpha} g_{\mu\nu}
\end{equation}
with
\begin{equation}
\alpha= -\frac{1}{2} \sqrt{\frac{d(d+2)}{3}}.
\end{equation}
In the rest of the paper, we identify the volume modulus as
\begin{equation}
{\cal V}=   \sigma^{-2\alpha}
\end{equation}
which has no kinetic terms. When $d=6$ we have $\alpha=-2$ and
\begin{equation}
{\cal V}=  \,\left(\frac{{V}}{{ V}_0}\right)^{2\over 3}
\label{defi}
\end{equation}
where ${V}$ is the volume of the extra dimensions.

 Defining $m_{\rm Pl}^2= 2  M_{2+d}^{d+2} { V}_0$, we can identify
 the action~(\ref{act}) with the supergravity action in the
 supergravity frame, following Chap.~31 of~\cite{Weinberg:2000cr},
\begin{equation}
S_4= \frac{m^2_{\rm Pl}}{2} \int d^4x \sqrt{- g}\left( f(T,\bar T)  R + 6f_{T\bar T} \partial_\mu T\partial^\mu \bar T\right)
\end{equation}
where $f(T,\bar T)= e^{-K(T,\bar T)/3m_{\rm Pl}^2}$ is the K\"ahler potential
\begin{equation}
K=-3 m_{\rm Pl}^2 \ln \frac{T+\bar T}{m_{\rm Pl}}.
\end{equation}
Notice that $f_{T\bar T}=0$ as $f(T,\bar T)= \frac{T+\bar T}{m_{\rm Pl}}$ is linear in the modulus $T$ such that
\begin{equation}
V= { V}_0 \,\left(\frac{T+\bar T}{m_{\rm Pl}}\right)^{3\over2}.
\end{equation}
We couple matter to the Jordan frame metric $g_{\mu\nu}$. When matter is supersymmetric defined by the superfield $C$, the scalar part reads
\begin{equation}
  S_4= \frac{m^2_{\rm Pl}}{2} \int d^4x \sqrt{- g} \left( f(T,\bar T)
  R + 6f_{T\bar T} \partial_\mu T\partial^\mu \bar T + 6 f_{C\bar C}  \partial_\mu  C \partial^\mu \bar C\right)
  \label{decoup}
\end{equation}
where $f_{C\bar C}= - \frac{1}{3m_{\rm Pl}^2}$
leading to
\begin{equation}
f(T,\bar T)= \frac{T+\bar T}{m_{\rm Pl}}- \frac{\vert C\vert^2}{3m^2_{\rm Pl}}
\end{equation}
corresponding to
\begin{equation}
K(T,\bar T)=-3 m_{\rm Pl}^2 \ln \left(\frac{T+\bar T}{m_{\rm Pl}}- \frac{\vert C\vert^2}{m^2_{\rm Pl}}\right)
\label{kah}
\end{equation}
Notice that in the supergravity frame defined by the metric $ g_{\mu\nu}$, the matter fields $C$ are canonically normalised corresponding to the Jordan frame for matter.
In the Einstein frame, the action reads
\begin{equation}
S_4= \int d^4x \sqrt{- g^E} \left(\frac{m^2_{\rm Pl}}{2}R^E - K_{i\bar \jmath} \partial_\mu \phi^i \partial^\mu \bar \phi^{\bar \jmath}\right)
\end{equation}
where $\phi^i=(T,C)$. The K\"ahler potential~(\ref{kah}) is the one
we use in this work.

The crucial ingredient that we will use in the following is that there exists a frame, here identified as the supergravity frame, where the volume modulus
has no kinetic terms. This is not special to the volume indeed. Indeed, using a Weyl transformation one can always remove the kinetic terms of one field at the price of having
a non-trivial rescaling of the metric. What is special about the volume modulus is that this Weyl transformation coincides with the change of metric which transforms the Einstein frame where the kinetic terms are non-vanishing to the Jordan frame where the kinetic terms vanish and the volume modulus decouples from matter, see~(\ref{decoup}) where the $C$ field does not couple to the volume modulus. Had it been otherwise, the construction presented below would be altered and integrating out the volume modulus leading to the identication with a $f({\cal R})$ theory would have been impossible.

\subsection{The volume modulus and  $f({\cal R})$ theories}\label{sec:VfR}

We are interested in the low-energy dynamics of the volume modulus obtained, for instance, after the compactification of
the  ten-dimensional effective field theory of type-IIA string theory down to four dimensions on a Calabi-Yau manifold of Hodge  numbers $h_{1,1}$ and $h_{2,1}$. This leads to
$h_{1,1}$ hypermultiplets and $h_{2,1}+1$ vector multiplets. The
hypermultiplets always contain the volume modulus; see
for instance~\cite{Antoniadis:2003sw}.  We will assume that the multi-modulus dynamics reduces, after integrating out all the extra scalars, to a  system
involving the volume modulus only. The reason for this hypothesis, as will be clear below, is that the volume modulus has the correct dimensionless coupling to matter fields $\beta= 1/\sqrt 6$, see~\cite{Brax:2017bcp,Brinkmann:2023eph} for instance. The effective dynamics of the volume modulus are then assumed to be determined by
a scalar potential $V_{\rm vac}({\cal V})$ determined by non-perturbative effects.
As shown in the
previous  section there is a frame where  the volume
modulus has no kinetic term. This frame is not the same as the string
or Einstein frame but a Jordan frame where matter couples to the
metric $g_{\mu\nu}$, see the previous section for details about the different frames.  After dimensional reduction,
the action  reads
\begin{equation}\label{e:L4d}
 S =\int d^4x \, \sqrt{-g} {\Mp^2\over2}
  \left( V_{\rm vac}({\cal V}) + {\cal V} {\mathcal R}\right)+ S_{\rm matter}(g_{\mu\nu},\psi),
\end{equation}
where ${\cal V}$ is the volume modulus and $S_{\rm
  matter}(g_{\mu\nu},\psi)$ is the matter action.
As mentioned above  the field ${\cal V}$ has no intrinsic dynamics as
no kinetic terms are present in the Jordan frame of Sect.~\ref{sec:dimred}.

Let  assume that there exists a function $f(\chi)$ of an auxiliary field $\chi$ such that
\begin{equation}\label{e:Vtof}
\chi=-\frac{dV_{\rm vac}({\cal V})}{d \cal V}, \qquad
{\cal V}_0 f(\chi)= V_{\rm vac}({\cal V})- {\cal V}\frac{dV_{\rm vac}({\cal V})}{d \cal V}.
\end{equation}
with ${\cal V}_0$ a given constant. Hence,  the function ${\cal V}_0f(\chi )$ is the Legendre transform of the potential
$V_{\rm vac}({\cal V})$. Under the  assumption that
$f''(\chi)\neq0$ one can identify the volume modulus as
\begin{equation}
  \frac{\cal V}{{\cal V}_0}= f'(\chi).
\label{inve}
\end{equation}
The gravitational part of the action~(\ref{e:L4d}) becomes
\begin{equation}
  {\cal S}_{g}=  {m^2_{\rm Pl}\over 2}\int d^4x \sqrt{-g} \left(f(\chi)-\chi f'(\chi)+ f'(\chi) {\cal
    R} \right)
\end{equation}
where the four-dimensional Planck mass
depends now on ${\cal V}_0$ as
\begin{equation}
  m^2_{\rm Pl}= {\cal V}_0 \Mp^2.
\end{equation}
One can then integrate out the $\chi$ field as it has no dynamics and
is a simple auxiliary field. An extremum of the resulting action
exists provided $f''(\chi)\ne 0$ which gives
\begin{equation}\label{e:chiR}
  \chi={\cal     R}
\end{equation} and as a result one obtains  that the
action~(\ref{e:L4d}) is a $f({\cal R})$ theory in the Jordan frame
\begin{equation}\label{e:fR}
S=  \frac{m^2_{\rm Pl}}{2}\int d^4x \sqrt {-g}\, f({\cal     R}) + S_{\rm matter}(g_{\mu\nu},\psi).
\end{equation}
We then have, under the assumption~\eqref{e:Vtof}, that the volume modulus dependence in~\eqref{e:L4d} is
equivalent to a $f({\cal R})$ theory.

\subsection{Volume stabilisation}\label{sec:Vstab}

{The effective potential $V_{\rm vac}({\cal V})$ follows from  non-perturbative effects which involve an intricate  interplay with the
dynamics of other moduli~\cite{Cicoli:2015wja,Brinkmann:2023eph}. In
the large volume limit corresponding to ${\cal V}\to \infty$ the
potential in the Einstein frame vanishes corresponding to the absence
of cosmological constant in ten dimensions. Here we will assume that the volume modulus is stabilised at a finite value ${\cal V}_0$ { in the supergravity frame} such that locally
in the vicinity of the minimum\footnote{The potential in the Einstein
  frame $V_E({\cal V})={\cal V}_0^2\, V_{\rm vac}({\cal V})/{\cal V}^2$ has a local minimum for ${\cal V}= {\cal V}_0 - 2V_{\rm vac}/(m^2 {\cal V}_0)$ which is close to ${\cal V}_0$ in the realistic cases discussed in Sec.~(\ref{sec:quant}). During inflation the Einstein potential is nearly constant and the end of inflation occurs around the local minimum. For an explicit example inspired from string theory, see Appendix~\ref{ap1}.}
\begin{equation}
V_{\rm vac}({\cal V})=- V_{\rm vac}-\frac{m^2}{2} ({\cal V}-{\cal V}_0)^2.
\label{quad}
\end{equation}
This is only valid close to the minimum. For large deviations from the
minimum, correction terms must be considered first and eventually, in
the large volume limit, this expression becomes not valid any more and should be replaced by  a function such that the potential in the Einstein frame $V_E({\cal V})= V_{\rm vac}({\cal V})/{\cal V }^2$ converges to zero for large ${\cal V}$. This regime is not the one we are interested in as we focus on the domain around the stabilised value ${\cal V}_0$ and will give sufficient conditions on the expansion of $V_{\rm vac}({\cal V})$ in its vicinity to guarantee that Starobinski's inflation is under control.

Using~(\ref{inve}),  we obtain
$\chi= m^2 ({{\cal V}}-{{\cal V}_0})$ and ${\cal V}_0f(\chi)=- V_{\rm vac}
\frac{m^2}{2}({\cal V}^2 -{\cal V}_0^2)$ leading to a $f({\cal R})$
action in  the Jordan frame
\begin{equation}\label{e:L4dstab}
  S= \int d^4x \sqrt{-g} \left( -\frac{V_{\rm vac}}{{\cal V}_0}+{\mpl^2\over2}{\cal R}+\delta c_0 {\cal
      R}^2\right)+ S_{\rm matter}(g,\psi)
\end{equation}
and the coefficient of the induced ${\cal R}^2$ is given by
\begin{equation}\label{e:deltac0}
\delta c_0= \frac{\mpl^2}{4m^2{\cal V}_0}
\end{equation}
depending on the mass term for the volume modulus. Hence, we find that a theory of the ${\cal R}^2$ type follows from stabilising the volume modulus
and restricting its scalar potential to a simple quadratic term. Moreover, the curvature of the scalar potential around the stabilised value ${\cal V}_0$ determines
the coefficient of the ${\cal R}^2$ term.
\subsection{The volume modulus as a scalaron}\label{sec:Vscal}
After having stabilised the volume modulus, the action
in~\eqref{e:L4dstab} is an $f({\cal R})$ action with
\begin{equation}
f({\cal R})= - 2\Lambda+ {\cal R} +\frac{{\cal R}^2}{2m^2{\cal V}_0 }.
\end{equation}
where $\Lambda=V_{\rm vac}/({\cal V}_0 m^2_{\rm Pl})$.
Performing the usual change of frame (see~\cite{Sotiriou:2008rp} for a
review)
\begin{equation}\label{e:Vphiscaling}
  g^E_{\mu\nu}=\frac{{\cal V}}{{\cal V}_0} g_{\mu\nu} = f'({\cal R}) g_{\mu\nu};  \qquad   \frac{{\cal V}}{{\cal V}_0}:=\exp\left(-2\beta \varphi \right);\qquad \beta={1\over \sqrt6}
\end{equation}
one gets in the Einstein frame
\begin{equation}
  S=\int d^4x \sqrt{-g^E} {\mpl^2\over2}\left({\cal R}_E-\frac12
    \partial_\mu \varphi\partial^\mu \varphi- V(\varphi)\right)+
  S_{\rm matter}\left( e^{2\beta\varphi}g^E_{\mu\nu},\psi\right)
\end{equation}
with
\begin{equation}
  V(\phi)=   {{\cal R} f'({\cal R})-f({\cal R})\over (f'({\cal
      R}))^2}.
\end{equation}
We see in this frame that the matter fields $\psi$ have the universal
coupling to the volume through the scaling factor
$e^{2\beta\phi/\mpl}$ with the strength controlled by $\beta=1/\sqrt6$.

The relation in~\eqref{e:Vphiscaling}
relates the volume ${\cal V}$ to the scalaron $\varphi$.
Under scale transformations $x^\mu \to x^\mu/\lambda$, the scalaron
field transforms as $\varphi\to \varphi -3/\beta \ln \lambda$, i.e.\
$\varphi$ is the Goldstone boson associated to the breaking of scale
invariance. As a result the evolution of $\varphi$ from $-\infty$ to
$\infty$ gives a ``reading'' in time of the renormalisation group
evolution from the Ultra-Violet to the Infra-Red.
In~\cite{Wetterich:1994bg} a cosmon field was considered from the
volume using~\eqref{e:Vphiscaling}. The difference here is that the scalar
field is the scalaron from a $f({\cal R})$ theory.

\subsection{The curvature expansion}\label{sec:highderivatives}

The action in~\eqref{e:L4d} is part of the low-energy effective action after compactification
and  can receive higher derivative corrections.
We now review the derivative expansion of the compactification to four dimensions
of the gravitational sector of  ten-dimensional string theory. The
discussion is given in the Jordan frame which is obtained by the Weyl
scaling from the string or Einstein  frame as  discussed in Sect.~\ref{sec:dimred}.

The curvature squared models are part of a four-dimensional string theory low-energy
effective action which has a derivative expansion of the type  in the Jordan frame
\begin{equation}\label{e:Leff4d}
 S^{\rm  eff}_{4d} =\int d^4x \, \sqrt{-g}
  \left({\Mp^2\over2}  \, {\cal V}_0 {\mathcal R}
    + c_0{\cal V}_0{{\mathcal R}^2}\right)+ \delta S^{\rm
    higher}_{4d}+S_{\rm matter}(g_{\mu\nu},\psi),
\end{equation}
where the higher derivative contribution is given by
\begin{equation}\label{e:highercorrections}
 \delta S_{4d}^{\rm high.} =\int d^4x \, \sqrt{-g}
  \Mp^2 \alpha_3\frac{{ R}^2}{M^2}\,\left( \tilde l_6^2 { R} +  {\cal V}_0\frac{\ell_s^4}{\tilde l_6^4}\sum_{p\ge 3} \frac{\alpha_{p+1}}{\alpha_3} (\ell_s^2 { R})^{p-1}  \right)  \,.
\end{equation}
Here we have assumed that the volume modulus is fixed at its value ${\cal V}_0$.
The  ${ R}^2$ couplings are modulus dependent and in the large
volume limit are dominated by the volume compactification (see~\cite{Gregori:1997hi,Harvey:1996ir} for
instance for a discussion of $R^2$ terms from string theory).
These expressions are symbolic, and ${ R}$ represents either  the Riemann tensor, the  Ricci tensor or the Ricci scalar.
We now briefly review the analysis of these higher derivative corrections as given in 
Appendix~A of~\cite{Burgess:2016owb} and
Sec.~5.1
of~\cite{Brax:2017bcp} (which we follow for the notations),
where it is argued that  the higher derivative corrections to the Einstein-Hilbert action are
suppressed in the large volume expansion.

The dimensionless coefficient $c_0$ of the scalar $\mathcal R^2$ term is given by
\begin{equation}\label{e:c0def}
c_0=  \alpha_3\frac{M_{\rm Pl}^2}{M^2}
\end{equation}
where the mass scales $\Mp^2$ and $M^2$ depend on the string coupling constant $g_s$, the
string scale $\ell_s$ and the compactification scale $\ell_6$
\begin{equation}\label{e:Masses}
  \Mp^2 \simeq \frac{l_6^6}{g_s^2\, \ell_s^8}\left(1+ \sum_{n\geq 3} \alpha_n
\frac{\ell_s^{2n}}{\tilde{l}_6^{\;2n}}\right); \qquad
M= \frac{\tilde l_6^2}{\ell_s^3}
\end{equation}
and the volume-dependent curvature  scale
\begin{equation}
\tilde l_6= {\cal V}_0^{1/6} l_6,
\end{equation}
which controls the typical averaged value of the ten-dimensional curvature $R$ over the
compactification manifold $M_6$
\begin{equation}
\int_{M_6} \frac{d^6 x}{V_6} \sqrt{-g} R^n  \sim \tilde{l}_6^{\;-2n} \, .
\end{equation}
A condition for the curvature expansion to make sense is that ${\tilde
  l}_6 \gg \ell_s$ where $\ell_s$ is the string scale.
The higher-order terms in the curvature expansion depend on the couplings
\begin{equation}
d_p \simeq \sum_{n\geq \max(p+1,3)} \alpha_n \left(\frac{\ell_s}{\tilde{l}_6}\right)^{2(n-p-1)} \, ,
\end{equation}
and the series are dominated by the first term in the regime ${\tilde l}_6 \gg \ell_s$.
This gives
\begin{equation}
d_0 \simeq \alpha_3 \left(\frac{\ell_s}{\tilde{l}_6}\right)^4 , \;\;\;
d_1 \simeq \alpha_3 \left(\frac{\ell_s}{\tilde{l}_6}\right)^2 , \;\;\;\;
p \geq 2: \;\; d_p \simeq \alpha_{p+1} .
\end{equation}

This immediately gives us indications that the ${\cal R}^2$ models are a valid description of the physics
at low-energy up to the string scale $M_s=1/\ell_s$ as long as
curvature is small in string units, i.e.\ $\ell_s^2 { R} \ll 1$. This suppresses all the higher curvature terms
and leaves only the dimension-four ${ R}^2$ terms and the dimension-six ${ R}^3$ corrections as
relevant at low energy. At the four derivative order, the Riemann
tensor square terms can be traded for a combination of the Ricci
scalar square and the Ricci tensor square
$({ R}_{\mu\nu}) ^2$ using the standard Gauss-Bonnet identity in
four dimensions. The $({ R}_{\mu\nu}) ^2$ term induces a massive
ghost spin-2 state, with a mass determined by the coefficient of the
coupling in the effective action~\cite{Stelle:1977ry}.
The coefficients of these terms are affected by the metric field
redefinition
$g_{\mu\nu}\to g_{\mu\nu}+\alpha{ R}_{\mu\nu} +\beta{\cal
  R}g_{\mu\nu}$ so that the coefficient
of the ${\cal R}^2$ is shifted by $(\alpha+3\beta)/3$ and the ${
  R}_{\mu\nu}^2$ term by $\alpha$~\cite{Tseytlin:1986zz,Forger:1996vj}.
By considering the dynamics of supersymmetry breaking involving the
volume modulus as in Sec.~5.2.2 of~\cite{Brax:2017bcp}, we can
consider  a situation where $\beta\gg\alpha$, and that the coefficient of
the ${ R}_{\mu\nu}^2$ term is  such that the mass of the massive spin-2
is  at the order of the UV cut-off of the theory. In fact, we require that the only physical compactifications are
the ones which lead to the absence of ghosts or such that the ghosts
are rejected at the UV cut-off scale of the theory. As a result,
the remaining higher derivative corrections in~\eqref{e:Leff4d}
are the powers of the Ricci scalar $\mathcal R$.

The ${\cal R}^2$ model only applies up to the curvature scale ${\cal R} \leq
\tilde{l}_6^{-2}$ which is typically much greater than practical
curvature scales. This suppresses the dimension-six  cubic ${\cal
  R}^3$ corrections compared to the
quadratic one, and we can focus on the ${\cal R}^2$ action
in~(\ref{e:Leff4d})
\begin{equation}\label{e:L4dR2}
   S^{\rm  eff}_{4d} =\int d^4x \, \sqrt{-g}
\left( {\Mp^2\over2} \left[ V_{\rm vac}({\cal V}) +{\cal V} {\mathcal R}\right]
    + c_0{\cal V}_0{{\mathcal R}^2}\right)+S_{\rm matter}(g_{\mu\nu},\psi).
\end{equation}
We can now apply the analysis of Sec.~\ref{sec:VfR} this time
starting with the action~\eqref{e:L4dR2}.   the resulting $f({\cal R})$ action in~\eqref{e:L4dstab}  after stabilisation of the
volume becomes
\begin{equation}\label{e:L4dstabR2}
  S^{\rm eff}_{4d}= \int d^4x \sqrt{-g} \left[{\mpl^2\over2} {\cal
      R}+\left(c_0  {\cal V}_0+\delta c_0\right) {\cal
      R}^2\right]+ S_{\rm matter}(g,\psi).
\end{equation}
The coefficient of the ${\cal R}^2$ is given by the  $c_0$ coefficient
in~\eqref{e:c0def} from the compactification of the string-induced
corrections and the corrections induced from the volume stabilisation
in~\eqref{e:deltac0}
\begin{equation}
  c_0{{\cal V}_0}+\delta c_0={\alpha_3 \Mp^2\over M^2} {{\cal V}_0}  +
  {\mpl^2\over 4m^2{\cal V}_0}= {\mpl^2\over 2}\left({\alpha_3\over
      M^2}+{1\over 2m^2{\cal V}_0}\right).
\end{equation}
Notice that both terms are large as $m\ll m_{\rm Pl}$ and
the compactification scale is such that  $\tilde l_6 \gg \ell_s$.
We can then just apply the formalism of Sec.~\ref{sec:Vscal} in the Jordan frame  with this time the function
\begin{equation}
  f({\cal R})={\cal R}+\left({1\over 2m^2{\cal V}_0}+{\alpha_3\over M^2}\right) {\cal R}^2.
\end{equation}
This allows us to identify the volume modulus as
\begin{equation}
{\cal V}= {\cal V}_0 + \left({1\over m^2}+{2\alpha_3{\cal V}_0\over M^2}\right ) {\cal R}.
\label{dev}
\end{equation}
This relation is not surprising as the volume modulus has no dynamics in the original Jordan frame and can therefore be integrated out exactly. As
the volume modulus mixes with the Ricci scalar of the string compactification, this implies that the volume modulus is stabilised up to fluctuations which
are parameterised by the Ricci scalar. As the Ricci scalar is also related to the scalaron, we find that the volume modulus and the scalaron become one and only one field.
We can also identify the scalaron used in the Weyl transformation to the Einstein frame as
\begin{equation}
\frac{{\cal V}}{{\cal V}_0}= e^{-2 \beta \varphi}.
\end{equation}
The volume modulus deviates from its stabilised value by an amount which is parameterised by  the scalaron. In the regime where the curvature is small compared to the large scale $M$, we find that such a deviation is small and the scalaron has a small excursion in Planck units.

In the following, we will reverse engineer the construction and obtain the low-energy $N=1$ supergravity description for a scalar field whose coupling to matter is given by $\beta=1/\sqrt{6}$. This will determine the K\"ahler potential of the associated chiral superfield. By imposing that the scalar potential is the one of the Starobinski model, we will determine the superpotential. The K\"ahler potential that we obtain is the one of the volume modulus with deformations parameterised by two parameters only. In a sense, this confirms that the effective $N=1$ supergravity description of Starobinski's model is related to the volume modulus and that its effective dynamics are determined by a superpotential whose shape is uniquely determined. Finding explicit string theory models whose behaviour mimics these results is a challenge left for future work.

\section{Supersymmetry}
\label{sec:susy}

\subsection{The volume modulus and its K\"ahler potential}

So far we have not taken into account one crucial ingredient:
supersymmetry. The compactifications that we consider should
result in a supergravity theory in four dimensions. Written in terms
of a modulus superfield $T$, its K\"ahler potential and its
superpotential $W(T,\bar T)$, the resulting $N=1$ supergravity theory should be
such that the matter fermions $\psi$ associated to another superfield
$C$ have a coupling given by $\beta=1/\sqrt6$. The archetypical example of this type of behaviour is given by the K\"ahler potential
\begin{equation}
K(T,\bar T)= -3 m_{\rm Pl}^2 \ln \left (\frac{T+\bar T}{m_{\rm Pl}} -
  \frac{C\bar C}{3m_{\rm Pl}^2}\right )\simeq  -3 m_{\rm Pl}^2 \ln
\frac{T+\bar T}{m_{\rm Pl}} + \mpl \frac{C\bar C}{T+\bar T}
\end{equation}
As we will recall below, this gives rise to the expected $\beta=\frac{1}{\sqrt 6}$ coupling.
Moreover, a constant superpotential
\begin{equation}
W= W_0
\end{equation}
gives rise to a vanishing scalar potential for the volume modulus.

In the following, we will generalise the K\"ahler potential and superpotential by imposing that the coupling to matter is determined by $\beta$ and that the scalar potential
follows from the ${\cal R}^2$ theory as expressed in the Einstein frame via the scalaron.

\subsection{From supergravity to the scalaron}

We define the scalaron field as the canonically normalised real part
of $T$, the universal K\"ahler modulus determining the volume of compactification, i.e.\ we require that
\begin{equation}
 d\varphi = -\,\sqrt{ 2 K_{T\bar T} (T,T) } {dT\over \mpl}
\label{norm}
\end{equation}
Notice the minus sign. Here the K\"ahler  potential is left unspecified and will be determined by requiring that matter couples to the scalarons like in $f({\cal R})$ theories.

Let us consider matter fields $C$ representing the superfield associated to the Weyl fermions $\psi_C$. We assume that the K\"ahler potential of these matter fields is in
\begin{equation}
K\supset \frac{C\bar C}{(T+\bar T)}.
\end{equation}
In supergravity this implies that the kinetic terms for the fermions are
\begin{equation}
i\frac{ \partial \bar \psi_C \sigma^\mu D_\mu \psi_C}{(T+\bar T)}
\end{equation}
which can be normalised according to
\begin{equation}
\psi= \frac{\psi_C}{(T+\bar T)^{1/2}}.
\end{equation}
Now a mass term for the fermions depends on the superpotential and the K\"ahler potential
\begin{equation}
{\cal L} \supset e^{K(T,\bar T)/2m_{\rm Pl}^2} \frac{\partial W}{\partial C^2} \psi_C^2
\end{equation}
corresponding to a Majorana mass term. Dirac mass terms can be constructed by taking pairs of superfield coupled to a Higgs field. In terms of the normalised fermions this becomes
\begin{equation}
{\cal L} \supset e^{K(T,\bar T)/2m_{\rm Pl}^2}(T+\bar T)  \frac{\partial W}{\partial C^2} \psi^2.
\end{equation}
As the superpotential cannot depend on the modulus field $T$ at the perturbative level, this mass term can be written as
\begin{equation}
{\cal L} \supset   A(\varphi) m \psi^2
\end{equation}
where $m= \frac{\partial W}{\partial C^2}$ is the modulus-independent mass and
\begin{equation}
A(\varphi)=  e^{K(T,\bar T)/2m_{\rm Pl}^2}(T+\bar T)\vert_{T(\varphi)}
\label{factor}
\end{equation}
is a factor dependent on the normalised field $\phi$.

\subsection{The K\"ahler potential from the $f({\cal R})$ matter coupling}\label{sec:kaehler}
We can try to find the K\"ahler potentials such that the normalised
real part of $T$, i.e.\ $\Re{e} (T)=\mpl\,t$, has a coupling given by
\begin{equation}\label{bingo}
A(\varphi)= e^{\beta \varphi} = e^{K(T,\bar T)\over 2m_{\rm
    Pl}^2} (T+\bar T)|_{T(\varphi)}.
\end{equation}
We set $K(T,\bar T)|_{T=\bar T=t} = k(t)
m_{\rm Pl}^2$. We find from
(\ref{norm}) that
\begin{equation}
  \left(d\varphi(t)\over dt\right)^2=\frac12 {d^2k(t)\over dt^2},
\end{equation}
where we used  that
\begin{equation}
{d^2k(t)\over dt^2}=4 K_{T\bar T}(T,\bar
T)|_{T=\bar T=t}\,.
\end{equation}
Imposing~(\ref{bingo})  gives the relation
\begin{equation}\label{e:ktophi}
  k(t)=2 \beta\varphi(t)-2\ln(2t)\,,
  \end{equation}
which leads to the differential equation
\begin{equation}
 \left(d\varphi(t)\over dt\right)^2= \beta
   {d^2\varphi(t)\over dt^2}+{1\over t^2}.
 \end{equation}
 We leave $\beta$ indeterminate, and we will show how $\beta$ is
uniquely determined to the value $\beta=1/\sqrt6$.
This equation has for solutions for all values of $a$ and $b$
\begin{equation}\label{pphi}
  \varphi(t)={a+\log(4)\over2\beta}+
  \frac12(-\beta+\sqrt{\beta^2+4})\ln(t)-\beta
  \ln\left( t^{ {\sqrt{\beta^2+4}\over\beta}}-b\right)\,.
\end{equation}
The K\"ahler potential reads (after redefining the arbitrary constant $a$)
\begin{equation}
  k(t)= a- (2+\beta^2- \beta\sqrt{\beta^2+4}) \log(t) -2\beta^2
  \log\left(t^{ {\sqrt{4+\beta^2}\over\beta}}-b\right)\,.
\end{equation}
Requesting  that $k(t) = - 3 \ln(2t)+ O(1)$ for large $t$ gives
\begin{equation}
  2+\beta^2+ \beta \sqrt{4+\beta^2}=3
\end{equation}
whose unique solution is  $\beta=1/\sqrt 6$.
Hence, we have found that the only coupling constant $\beta$ in the identification between the coupling of fermion fields in supergravity and
a scalaron coupling is the one compatible with the $f({\cal R})$ structure.

The resulting K\"ahler potential is given by
\begin{equation}\label{e:ksol}
  k(t)= a-{4\over 3}\ln(t)-{1\over 3}\ln( t^5-b) =a - 3 \ln t +\frac{1}{3} \sum_{n\geq
  1}\frac{b^n}{t^{5n}}.
\end{equation}
The associated  scalaron solution reads
\begin{equation}
    2\beta \varphi(t)=a+\log(4)+{2\over 3}\log(t)-{1\over 3}\log( t^5-b)= a+\log(4)-\log t+\frac13\sum_{n\geq1} {b^n \over t^{5n}}.
\end{equation}
When  $b<0$ we have that $\varphi(t)\to-\infty$ for $t\to 0$ and
  $t\to+\infty$ with $\varphi(t_*)=0$ for
  \begin{equation}
    64 e^{3a}  t_*^2= t_*^5-b    \,.
  \end{equation}
 When $b>0$ we must have $t > b^{-1/5}$, and   $\varphi(t)\to+\infty$ for $t\to 1/b^{1/5}$
  and $t\to +\infty$.

In conclusion, the most general K\"ahler potential which is compatible with the coupling to matter of $f({\cal R})$ models is a modification of the
no-scale models by a series of corrections in $1/t^5$.
\subsection{The scalaron potential and the reconstruction of the superpotential}

The ${\cal R}^2$ model is associated to a specific scalar potential for the scalaron $V(\varphi)$. This is then equivalent to having a scalar potential $V(\varphi(t))$ using the mapping $\varphi(t)$.
In the following we will focus on the theory
\begin{equation}
S= \frac{m^2_{\rm Pl}}{2} \int d^4x f({\cal R})
\end{equation}
where we introduce an explicit cosmological constant
\begin{equation}
f({\cal R})=-2\Lambda+ {\cal R} + c {\cal R}^2.
\end{equation}
There the coefficient $c$ is related to the coefficient $c_0$ and the
correction $\delta c_0$ discussed previously by
\begin{equation}
c_0+\delta c_0= \frac{m^2_{\rm Pl} c}{2}.
\end{equation}
We have then the identification
\begin{equation}
  e^{-2\beta\varphi}=1+2c {\cal R}
\end{equation}
and finally the scalar potential
\begin{equation}\label{e:Vphi}
  V(\varphi)= {\left(1-e^{2\beta\varphi}\right)^2\over 4c} +2\Lambda e^{4\beta\varphi},
\end{equation}
which becomes a function of $t$ using~(\ref{pphi})
\begin{equation}\label{e:V}
  V(t)=\frac{\left(32 \Lambda  c +8\right) t^{\frac{4}{3}} \, e^{2a}}{\left(t^{5}-b \right)^{\frac{2}{3}} c}-\frac{4 t^{\frac{2}{3}}e^{a}}{\left(t^{5}-b \right)^{\frac{1}{3}} c}+\frac{1}{2 c}.
\end{equation}
Interestingly, this potential has a minimum for the value
\begin{equation}
e^{2\varphi_{\rm min}}= \frac{1}{1+ 8 c \Lambda}
\end{equation}
for which the potential energy becomes
\begin{equation}
V(\varphi_{\rm min})= \frac{2\Lambda }{1+ 8 c\Lambda}.
\end{equation}
When $\Lambda >0$, this implies that supersymmetry is broken at the minimum of the potential. We will concentrate on this case below
and therefore supersymmetry will always be spontaneously broken in the supergravity models that we will consider.

Let us recall that in $N=1$ supergravity and using Planck units in the following equations (the Planck scale can be easily reinstated by simple
dimensional analysis), the scalar potential reads
\begin{equation}
V(t)= e^{k(t)}\left( k^{T\bar T}\vert \partial_T W + \partial_T k W\vert^2 -3 \vert W\vert^2\right),
\label{superpot}
\end{equation}
where $k^{T\bar T}= 1/\partial_T \partial_{\bar T} k$. We have also $k_{T\bar T}= \frac{k''}{4}$ where $k''=\frac{d^2k}{dt^2}$.
In the following we will focus on a real superpotentials when $T=\bar
T=t$ is real, since  are focusing on the volume dependence which is the real part of
the $T$ modulus.  We can always extend this  by analytic continuation (in an open subset of the complex plane $\mathbb C$ containing the
real line $\mathbb R$) to  an holomorphic superpotential
$\mathfrak w(T)$
matching  $w(t)$ on the real line.  Equation~(\ref{superpot}) becomes
\begin{equation}\label{e:W}
\left(\frac{d W(t)}{dt}+\frac12 \frac{dk(t)}{dt} W(t)\right)^2=\frac{ k''(t)}{4}\left(3 W(t)^2+ e^{-k(t)}
  V(t)\right).
\end{equation}
It is convenient to define the function
\begin{equation}
w(t)= W(t) e^{k(t)\over2}
\end{equation}
which is nothing but $ w(t) = e^{G(t)/2}$ where $G(t)= k(t)+ \ln \vert W(t)\vert^2$ characterises the full $N=1$ Lagrangian and shows the underlying K\"ahler invariance
$k(T)\to k(T)+ f(T)+\bar f(\bar T)$, $ W(T)\to e^{-f(T)}W(T)$ where $f$ is
a holomorphic function of $T$. Using this function we have
\begin{equation}\label{e:w}
\left(\frac{dw(t)}{dt}\right)^2-\frac34 k''(t)  w(t)^2=\frac{ k''(t)}{4} V(t).
\end{equation}

Given the function $V(t)$ obtained from the $f({\cal R})$ theory, we can solve this differential equation and find the most general superpotential compatible with the ${\cal R}^2$ structure.

\subsection{The superpotential associated to ${\cal R}^2$}

As a sanity check,  let us consider a no-scale model with $V(t)=0$ and
$k(t)=-3 \ln t$, i.e.\ $b=0$ in~(\ref{e:ksol}). From the
differential equation~\eqref{e:W}
we retrieve the familiar no-scale model where $W(t)=w_0$ is a constant
and another solution $W(t)=w_0 \, t^{3}$ for which the potential also
vanishes.

For the  superpotential associated to ${\cal R}^2$ with $k(t)$ given
in~(\ref{e:ksol}), and the potential~(\ref{e:V}), the differential
equation has the unique solution (this is unique up to an overall sign)
\begin{multline}
  w_\pm(t)={\pm1\over\sqrt{-6c}}\Big(1-\frac{4 \,{\mathrm
      e}^{a}}{t}+\frac{32 \left(9 \Lambda  c +1\right) {\mathrm e}^{2
      a}}{9 t^{2}}+\frac{128 \left(9 \Lambda  c +1\right) {\mathrm
      e}^{3 a}}{81 t^{3}}\cr+\frac{\left(290304 (\Lambda
      c)^{2}+50688 \Lambda  c +2048\right) {\mathrm e}^{4 a}}{729
    t^{4}}+\frac{\left(-3151872 (\Lambda c)^{2}-313344 \Lambda  c
      +4096\right) {\mathrm e}^{5 a}}{6561 t^{5}}\cr
  +{1\over t^6} \, \left(\frac{\left(213663744 (\Lambda c)^{3}+78446592 (\Lambda c)^{2}+7626752 \Lambda  c +172032\right) {\mathrm e}^{6 a}}{6561}-\frac{4 b \,{\mathrm e}^{a}}{3}\right)+O\! \left(\frac{1}{t^{7}}\right)
\Big).
\end{multline}
The superpotential is obtained by $W_\pm(t)= w_\pm(t)
\exp(-k(t)/2)$.  This is uniquely determined from the K\"ahler potential
and the $f({\cal R})$ functional.

Since $k(t)\simeq -3\log(t)$ for large $t$ we get a diverging
expression at large volume for the superpotential. This large $t$
behaviour can be removed  using a K\"ahler transformation (see
Chap.~XXIII in~\cite{WessBagger})
  $K(T,\bar T)\to K(T,\bar T)+f(T)+\bar f(\bar T)$ and $W(T,\bar T)\to
  e^{-f(T)} W(T,\bar T)$ with $f(T)={3\over 4} \log(T/\mpl)+\log(\pm i)$. This shifts the K\"ahler potential to
  \begin{equation}
    K(T,\bar T)= m_{\rm Pl}^2 \left(a - \frac43 \log\left(T+\bar
        T\over \mpl\right)-\frac13
    \log\left(\left(T+\bar T\over \mpl\right)^5-b\right)+\frac34\log\left(T\bar T\over\mpl^2\right)\right)
  \end{equation}
  with
  \begin{multline}
  W(t)
=\mpl^3 { e^{-{a\over2}}\over\sqrt{6 c}} \,
                               \Big(1-\frac{4 \,e^{a}}{ t}+\frac{32
                               \left(9 \Lambda  c +1\right) e^{2 a}}{9
                             t^{2}}+\frac{128 \left(9 \Lambda  c
                               +1\right) e^{3 a}}{81\, t^{3}}\cr
                               +\frac{512 \,e^{4 a} \left(567
                                   (\Lambda c)^{2}+99 \Lambda  c
                                   +4\right)}{729 \,
                                 t^{4}}+{1\over t^5}\left(\frac{\left(-6303744
                                     (\Lambda c)^{2}-626688 \Lambda
                                     c +8192\right) e^{5
                                     a}}{13122}-\frac{b}{6}\right)
+\cdots\Big)   .
\end{multline}
The K\"ahler potential is lacunar and the superpotential is an
infinite series in $1/t$ where $\Re e(T)=\mpl \, t$.
This is valid when $b\ne 0$ where the first correction in $b$ appears at the $t^{-5}$ order.

When $b=0$, the K\"ahler potential is $k(t)=a-3\log(t)$, the $f({\cal
  R})$ potential is
\begin{equation}
V(t)|_{b=0}=  \frac{\left(32 \Lambda  c +8\right) {\mathrm e}^{2 a}}{t^{2} c}-\frac{4 \,{\mathrm e}^{a}}{t c}+\frac{1}{2 c}
\end{equation}
and the superpotential is given by
  \begin{multline}
  W(t)|_{b=0}
=\mpl^3 { e^{-{a\over2}}\over\sqrt{6 c}} \,
                               \Big(1-\frac{4 \,e^{a}}{t}+\frac{32
                                 \left(9 \Lambda  c +1\right) e^{2
                                   a}}{9 t^{2}}+\frac{128 \left(9
                                   \Lambda  c +1\right) e^{3 a}}{81
                                 t^{3}}\cr+\frac{\left(290304
                                   (\Lambda  c)^{2}+50688 \Lambda  c
                                   +2048\right) e^{4 a}}{729
                                 t^{4}}-\frac{2048 \,
                                 \left(1539 (\Lambda c)^2+153
                                   \Lambda  c -2\right) e^{5 a}}{6561
                                 t^{5}}\cr+\frac{\left(213663744
                                  ( \Lambda c)^{3}+78446592 (\Lambda c)^{2}+7626752 \Lambda  c +172032\right) e^{6 a}}{6561 t^{6}}
+\cdots\Big)   .
\label{WW}
\end{multline}
We find an expansion in $t e^{-a}$ where
$a$ is the constant of integration in the K\"ahler potential. Actually
shifting the value of $a$ amounts to redefining the Planck mass $\mpl$.

We notice that the  cosmological constant $\Lambda$
and the coefficient $c$ of the ${\cal R}^2$ term appear in the
combination $\Lambda c$.
We now turn to the quantum evolution of $c$, from the Ultra-Violet (the inflationary
period)  to the Infra-Red regime (the late time behaviour).  We will
relate the evolution of $c$ along the renormalisation group flow to
the values of the evolution of the cosmological constant.

\section{The quantum evolution of $c$}
\label{sec:quant}

The model that we have considered so far is valid in the Ultra-Violet at high energy, i.e.\ at the compactification scale. We will recall how inflation appears in this setting and then
link this behaviour to the properties of the scalaron in the Infra-Red regime.

\subsection{Starobinski inflation}

The ${\cal R}^2$ model  is a very good candidate for inflation. During the inflationary regime the potential in dimensionful units, defined as $V_{\rm inflation}(\varphi)=m_{\rm Pl}^2 V(\varphi)/2$,   is
\begin{equation}
V_{\rm inflation}(\varphi)= \frac{m^2_{\rm Pl}}{8c}\left(e^{2\beta \varphi} -1\right)^2 + \Lambda m^2_{\rm Pl} e^{4\beta \varphi}.
\end{equation}
Inflation takes place in a  quasi-de Sitter region of field space with $\varphi \to -\infty$ such that $e^{2\beta \varphi}\ll 1$. This is the large volume regime where $t\to \infty$. Notice that the cosmological constant $\Lambda$ plays no role in this regime. We have therefore
\begin{equation}
V_{\rm inflation}(\varphi)\simeq  \frac{m^2_{\rm Pl}}{8c}\left(1- 2e^{2\beta \varphi}\right).
\end{equation}
The slow roll parameters are then
\begin{equation}
  \epsilon= \frac{1}{2} \left(\frac{V_{\rm inflation}'}{V_{\rm inflation}}\right)^2
   \simeq {8\beta^2} e^{4\beta \varphi},
\end{equation}
and
\begin{equation}
\eta=  \frac{V_{\rm inflation}''}{V_{\rm inflation}} \simeq -8\beta^2 e^{2\beta \varphi}.
\end{equation}
Observable scales by the cosmic microwave background (CMB) are $N$ $e$-folding before the end of inflation where $\epsilon_{\rm end}=1$.
For these scales, the $\eta$ parameter dominates over  $\epsilon$
\begin{equation}
\epsilon_\star \ll \vert \eta_\star\vert,
\end{equation}
and the spectral index is then given by
\begin{equation}\label{e:ns}
n_s-1= 2 \eta_\star \simeq -16\beta^2 e^{2\beta \varphi_\star}.
\end{equation}
This determines $\varphi_\star$ as $n_s-1\simeq -0.0351$ according to Planck~\cite{Planck:2018vyg}.
The number of $e$-folding is $a_{\rm end}/a_\star= e^N$ where $a$ is the scale factor of the Universe with
\begin{equation}
N= \int_{\varphi_\star}^{\varphi_{\rm end}} \frac{V_{\rm inflation}}{V_{\rm inflation}'} d\varphi.
\end{equation}
We find
\begin{equation}
N\simeq \frac{1}{8\beta^2} e^{-2\beta \varphi_\star},
\end{equation}
and therefore
\begin{equation}
n_s-1= -\frac{2}{N},
\end{equation}
which determines $N\simeq 56$ as $n_s-1\simeq -0.0351$.  At the end of inflation we have
\begin{equation}
e^{-2\beta \varphi_{\rm end}}=2\sqrt 2\beta,
\end{equation}
where $\epsilon (\varphi_{\rm end})=1$. Notice that the approximation $\varphi\to -\infty$ is not really valid toward the end of inflation. In principle this means that numerics is required to evaluate when $\epsilon=1$ and inflation stops.
After the end of inflation the field oscillates around zero and eventually reaches $\varphi \ll 1$.

\subsection{Validity of the curvature expansion}

At the end of inflation, the field $\varphi$ settles at the minimum of the potential for $\varphi \ll 1$ where it becomes massive with
\begin{equation}\label{e:massphi}
m^2_\varphi= \frac{ \beta^2}{c} + 16 \beta^2 \Lambda,
\end{equation}
where the second term is negligible at the end of inflation.
The value of $c$ at the end of inflation can be deduced from the normalisation of the CMB spectrum as
\begin{equation}
 \frac{V_{\rm inflation}(\varphi_\star)^3}{m_{\rm Pl}^6(V'_{\rm inflation}(\varphi_\star))^2}\simeq 2\cdot 10^{-11}
\end{equation}
evaluated at the value of $\varphi_\star$ determined in~\eqref{e:ns},
which implies that
\begin{equation}
c^{\rm end}m_{\rm Pl}^2 \simeq {5\over8}\cdot 10^{10}
\beta^2 N^2.
\label{cinf}
\end{equation}
This is the value of $c$ at the end of inflation.
This gives a mass for $m_\varphi$ around $2\cdot 10^{-7} m_{\rm Pl}$.

Now we can come back to the curvature expansion and check that the expansion in powers of the Ricci scalar is valid.
The expansion is valid as long as $\tilde l_6^2 {\cal R}\ll 1$ and $ l_s^2 {\cal R}\ll 1$. During inflation we have $2c{\cal R}\approx \exp(-2\beta\varphi)$ and during inflation
\begin{equation}
2\sqrt 2 \beta \le e^{-2\beta\varphi} \le 8\beta^2 N
\label{excu}
\end{equation}
implying that
\begin{equation}
\tilde l_6^2 \ll \frac{c}{4\beta^2 N}.
\end{equation}
As $c$ is given by~(\ref{cinf}) we find that
\begin{equation}
\tilde l_6 \le \mpl^{-1} \sqrt{\frac{5N}{32}} 10^5
\end{equation}
implying that the compactification scale must be
\begin{equation}
\tilde l_6^{-1}\ge 3.4\times  10^{-6} \mpl,
\end{equation}
which is close to the Grand Unified Theory scale. This confirms that the curvature expansion in~\eqref{e:highercorrections} is valid.

\subsection{Validity of the quadratic expansion for the scalar potential}
\label{sec:val}
 We have shown that the quadratic terms around the minimum value ${\cal V}_0$ of the non-perturbative scalar potential
$V_{\rm vac}({\cal V})$ as a function of the volume modulus ${\cal V}$
in~(\ref{quad}) are enough to generate the inflationary Starobinski potential. One issue with this description is that the excursion of the volume modulus~(\ref{excu}) during inflation must be large and could jeopardise the quadratic approximation. Using~(\ref{defi}) we have
\begin{equation}
\frac{V}{V_6}= \left(\frac{\cal V}{{\cal V}_0}\right)^{3/2}
\end{equation}
where $V_6=\ell_6^6$ is the stabilised volume of the compactification manifold. Using dimensional analysis, we can write the non-perturbative potential
as
\begin{equation}
V_{\rm vac}({\cal V})= M^2 F(M^6V )= M^2 F\left( M^6 V_6 \left(\frac{\cal V}{{\cal V}_0}\right)^{3/2}\right)= M^2 G\left(\alpha \frac{\cal V}{{\cal V}_0}\right)
\end{equation}
where $G(x)\equiv F(x^{3/2})$ is a non-perturbative function which is assumed to vanish when ${\cal V}\to \infty$. Here we have used, in the spirit of effective field theories, that the low-energy dynamics
of the volume modulus are determined by a single non-perturbative scale $M$ and the function $G(x)$. We have introduced the dimensionless parameter $\alpha= (M^6 V_6)^{2/3}$.
Now the function $-G$ is assumed to have minimum for $x=\alpha$, so we can expand in Taylor series
\begin{equation}
V_{\rm vac}({\cal V})= M^2 G(\alpha)+ \frac{g_2}{2} \alpha^2 M^2  \left(\frac{\cal V}{{\cal V}_0}-1\right)^2 + M^2 \sum_{n\ge 3} \frac{g_n}{n!} \alpha^n \left(\frac{\cal V}{{\cal V}_0}-1\right)^n
\end{equation}
where we assume that $g_n={\cal O}(1)$. The first terms lead to  the quadratic Lagrangian~(\ref{quad}) with
\begin{equation}
V_{\rm vac}= -M^2 G(\alpha), \ m^2= -g_2 \alpha^2 M^2
\end{equation}
whilst the higher-order terms are negligible as long as
\begin{equation}
M \ell_6 \ll (8\beta^2 N)^{1/4}
\end{equation}
corresponding to a small suppression of the non-perturbative scale $M$ compared to the compactification scale $\ell_6^{-1}$.

\subsection{Post-inflationary era}

 After inflation which occurs at high energy, we follow a
  Wilsonian approach, reviewed in~\cite{Georgi:1993mps,Manohar:2018aog,Neubert:2019mrz,Burgess:2020tbq} for instance, and consider the effective action obtained by integrating out all the momentum scales larger than
a given scale $\mu$, $\mu\ll |p|$ and much smaller than the compactification scale,
to obtain the Wilson effective action at leading
order
\begin{equation}
S_W= \frac{m^2_{\rm Pl}}{2} \int d^4x f({\cal R};\mu)
\label{action1}
\end{equation}
the Wilsonian Lagrangian reads
\begin{equation}
f({\cal R})=-2\Lambda(\mu)+ {\cal R} + c(\mu) {\cal R}^2.
\label{action2}
\end{equation}
 Having fixed the Planck mass $m^2_{\rm Pl}$, which is then
  independent of $\mu$ and fixes the scales, the parameters in the
  Wilson effective action $S_W$ are $\mu$ dependent.  The dependence
  of the Lagrangian parameters in~(\ref{action1}) and~(\ref{action2})
  is given by the renormalisation group equation that we will present
  below and review briefly in Appendix~\ref{app:reg}. The initial
  values of the renormalisation group are taken at the end of
  inflation corresponding to the reheat temperature $T_{\rm reh}$
  taken to be larger than any physical masses of the particles in the
  spectrum of the theory.  During the cosmological evolution, the change of the scalaron mass and
  couplings follow the renormalisation group flow with respect to the
  scale $\mu$ which is cosmological time dependent, i.e. should be adapted to describe each cosmological era.

Although the Wilson effective action depends on the scale
  $\mu$, the physical observables are independent of that scale. In
  our case the scale-independent observables are  the energy density
  $\rho_{\rm vac}$ in~\eqref{e:rho-vac-me}  and the physical scalar mass $m_\varphi$ obtained  both in the IR corresponding to the limit $\mu \to 0$. In particular, the physical scalaron mass sets the range of the new scalar interaction to matter which can be tested using experiments such as E\"ot-Wash~\cite{Lee:2020zjt}.

The couplings evolve each time a particle species of mass $m$ is integrated out~\cite{Georgi:1993mps}, i.e. when $\mu \le m$. In the history of the Universe, particles in the thermal bath are integrated out when the temperature falls below the mass $m$. This allows us to estimate $\mu \sim T$. Let us review below how the renormalisation evolution of the vacuum energy and the coupling $c$ can be inferred. For explicit details on the regularisation procedure, see Appendix~\ref{app:reg}.}
Let us consider the quantum corrections to $m_\varphi$. As $\varphi$ couples to fermions like $\frac{\beta m_\psi }{m_{\rm Pl}} \varphi \bar \psi \psi$, the one-loop contribution to the effective mass of $\varphi$ is
\begin{equation}
\delta \bar m^2_\varphi(\mu)= -\frac{\beta^2 m_\psi^2 }{m_{\rm Pl}^2} \int \frac{d^4 p}{(2\pi)^4} \frac{1}{p^2 + m_\psi^2}.
\end{equation}
The divergent integral has to be renormalised. Using dimensional regularisation and the decoupling scheme where the effects of the corrections are only non-vanishing at the renormalisation scale $\mu \ge m_\psi$, we have
\begin{equation}
\delta \bar m^2_{\varphi}(\mu) =-\frac{\beta^2 m_\psi^4 }{16 \pi^2 m_{\rm Pl}^2} \ln \frac{\mu^2}{m^2_\psi} \theta (\mu -m_\psi),
\end{equation}
where we have introduced the Heaviside function $\theta(x)=1$ for $x>0$ and 0 otherwise.
This implies that the quantum corrections reduce the mass of the
scalar when $\mu$ decreases and, therefore, increase the value of $c(\mu)$ from~\eqref{e:massphi}.
In fact the scalaron couples to both massive scalars and massive vector
bosons implying that the correction to the mass at one-loop order is
given by the supertrace
\begin{equation}
\delta \bar m^2_{\varphi}(\mu) =-\frac{\beta^2  }{16 \pi^2 m_{\rm Pl}^2}{\cal
  S}tr\left( M^4  \ln \frac{\mu^2}{M^2}\, \theta(\mu-M)\right),
\end{equation}
where $M$ is the mass matrix of all the massive particles with a mass less than $\mu$.
This corresponds to a renormalisation group equation (see Appendix~\ref{app:reg} for a physical discussion about the regularisation method):
\begin{equation}
\frac{d \bar m^2_\varphi(\mu)}{d\ln \mu}= -\frac{\beta^2  }{8 \pi^2 m_{\rm Pl}^2}{\cal S}tr( M^4) \theta(\mu-M).
\end{equation}
Using the renormalisation group equation for the cosmological constant
\begin{equation}
\frac{d\Lambda(\mu)}{d\ln \mu}= -\frac{ 1 }{32 \pi^2 m_{\rm Pl}^2}{\cal S}tr( M^4)\theta(\mu-M),
\end{equation}
where again only particles with  a mass less than $\mu$ contribute. Combining these expressions using~\eqref{e:massphi} 
we get the renormalisation group equation for $c(\mu)^{-1}$ with $\beta=1/\sqrt6$
\begin{equation}
\frac{d c(\mu)^{-1}}{d\ln \mu}= \frac{7  }{16 \pi^2 m_{\rm Pl}^2}{\cal S}tr( M^4)\equiv -{14} \frac{d\Lambda(\mu)}{d\ln \mu}.
\end{equation}
This implies that
\begin{equation}
c_{\rm {IR}}^{-1}= c_{\rm {end }}^{-1} +{7} \left({\Lambda}_{\rm end} -  {\Lambda_{\rm IR}}\right) -{14} \sum_i \Lambda_i,
\end{equation}
where we have introduced the sum over the jumps of the cosmological
jumps when a phase transition happens corresponding to a jump of the
value of the cosmological constant by $\Lambda_i$. This can be complemented with the evolution of the cosmological constant to the deep IR where $\mu$ is much lower than all the particle masses and give
\begin{equation}
{\Lambda_{\rm IR}}={\Lambda}_{\rm end}+  \sum_i \Lambda_i
\end{equation}
 where $\Lambda_{\rm IR}$ is the cosmological constant obtained
  after integrating out all the quantum fluctuations and sending $\mu\to 0$ in the Wilson effective action. This corresponds to the cosmological constant in the full 1PI effective action of the theory.
We then obtain our final relation
\begin{equation}
c_{\rm {IR}}^{-1}= c_{\rm {end }}^{-1} -{7} \left({\Lambda}_{\rm end} -  {\Lambda_{\rm IR}}\right).
\label{balance}
\end{equation}
We have used $1+\frac{1}{\beta^2}=7$.
Notice that a positive cosmological constant at the end of inflation
would naturally lead to a smaller value of $c$ in the IR, i.e.\ an
increase in  the Lagrangian effective mass of the scalaron. We remark, as well, that the influence of the higher derivative terms that arise from the
non-renormalisability of the theory are negligible in the IR.
As the cosmological constant $\Lambda_{\rm end}$ at the end of inflation are not directly determined by the experimental data, i.e.\ the dynamics of the Starobinski model
is not influenced by the cosmological constant which only plays a role
toward the end of inflation, we cannot calculate
$c_{\rm IR}$ by following  the renormalisation group evolution from
the UV to the IR. We have to resort in Sec.~\ref{sec:stability} to low-energy stability arguments to bound the value of $c_{\rm IR}$.

Finally let us comment on supersymmetry breaking in these models. Supersymmetry is broken dynamically during inflation and then from the end of inflaton onward once the scalaron settles at its minimum, the vacuum energy corresponding to the minimum of the scalar potential becomes
\begin{equation}
V_{\rm min}= \frac{\Lambda m_{\rm Pl}^2}{1+ 8\Lambda c}
\end{equation}
which evolves with the renormalisation flow. In the IR, this coincides with
\begin{equation}
V_{\rm min} \simeq  \Lambda_{\rm IR} m_{\rm Pl}^2
\end{equation}
as we will see below that $ c_{\rm IR} \Lambda_{\rm IR} \ll 1$. Hence, at low energy supersymmetry is spontaneously broken, although extremely softly, by the small vacuum energy of the Universe.

\subsection{Low energy stability}\label{sec:stability}

The value of $c$ in the IR cannot be directly deduced from its value at the end of inflation without a detailed knowledge of high-energy physics and all the phase transitions between inflation and the present Universe.

On the other hand, the value of $c$ in the Infra-Red regime can be
bounded by phenomenological stability
arguments~\cite{Brax:2019iut}. We will assume that the scalaron
becomes light in the IR with  the physical mass $m_\varphi$, given by the value of effective mass $\bar m_\varphi(\mu)$ for $\mu=0$, which is directly
related to $c_{\rm IR}$  by the relation in~\eqref{e:massphi}. In practice, we will take the mass of the
scalaron much smaller than the electron and neutrino masses. In this
model the scalaron at low energy is assumed to be the lightest
massive particle in the Universe.  Defining $\rho_\Lambda= \Lambda m^2_{\rm Pl}/2$, the renormalisation group evolution in the deep IR below all particle masses gives
\begin{equation}
\rho_{\rm vac}= \rho_{\rm \Lambda}(m_e) + \frac{m_\varphi^4}{64\pi^2} \ln
\frac{m_e^2}{m_\varphi^2} - 2\sum_{f=1}^3
\frac{m_f^4}{64\pi^2} \ln \frac{m_e^2}{m_f^2} .
\label{e:rho-vac-me}
\end{equation}
where the observational value of the vacuum energy is simply
$
\rho_{\rm vac} \simeq 2.7 \times
10^{-11}~\textrm{eV}^4.
$
The vacuum energy at the energy of the electron mass has been denoted by $\rho_{\rm \Lambda}(m_e)$. This encapsulates our lack  of knowledge of the physics at scales larger than $m_e$.
 The neutrinos also contribute at low energy, and we have for the two possible hierarchies of neutrino masses.
For both ordering the neutrino contribution is bounded (see Sec.~V
of~\cite{Brax:2019iut})
\begin{equation}\label{e:NeutrinoBounds}
2\times 10^4\,\rho_{\rm vac} \leq \sum_{f=1}^3  {m_f^4\over
  64\pi^2}\,\log\left(m_e^2\over m_f^2\right)\leq 2\times
10^5\rho_{\rm vac}\,.
\end{equation}
This is obviously a large contribution which exemplifies the nature of the cosmological constant problem even at low energy.

Let us now invoke the stability argument of~\cite{Brax:2019iut}. The vacuum energy $\rho_{\Lambda}(\mu)$ cannot be too negative; otherwise,
any bound structure in the Universe whose constituents have a typical energy $\mu$ would collapse faster than the age of the Universe. We will therefore impose
that $\rho_\Lambda (\mu) \ge -\rho_\Lambda$ to guarantee the stability of the Universe.

The X-ray emitting gas of a galaxy cluster has
a typical temperature of $T_{\rm X} \sim 1$ keV. These systems typically appeared at a redshift $z \gtrsim 0.1$
and already have a lifetime of the order of the age of the Universe. At these energies, the vacuum energy corresponds $\rho_\Lambda (m_e)$ and the absence of collapse
of the clusters over the age of the Universe implies that $\rho_\Lambda(m_e) \gtrsim -\rho_{\rm vac}$. This implies that~\cite{Planck:2018vyg,Giusarma:2016phn,Vagnozzi:2017ovm,Giusarma:2018jei,DiValentino:2021imh,Esteban:2016qun}
\begin{equation}
m_\varphi\lesssim \bar m_\nu=(m_1^4+m_2^4+m_3^4)^{1\over4}\simeq 0.1~\textrm{eV} .
 \label{m-mnu-upper}
\end{equation}
 Similarly, the scalaron could appear as contributing to a fifth force in gravitational experiments~\cite{Stelle:1977ry}
  \begin{equation}\label{e:Vpot}
    V(r) = -{G_N M\over r}\, \left( 1+{1\over3}e^{-m_\varphi r}\right)    \,.
  \end{equation}
  The absence of evidence for  short-range forces  in the
E\"ot-Wash experiment~\cite{Hoyle:2004cw,Kapner:2006si,Lee:2020zjt} provides an upper bound on the range of scalar forces $d\le 52~\mu{\rm m}$ corresponding to the strong lower
bound
\begin{equation}
m_\varphi\gtrsim 3.8\times 10^{-3}~{\rm eV}\,.
\end{equation}
We have thus an interval of masses for the nearly massless scalaron. This is a fairly narrow interval provided the scalaron is less massive than the electron mass.
This implies in particular
\begin{equation}
  \Lambda_{\rm IR} c_{\rm  IR}^2\simeq 8\pi G_N
\end{equation}
a relation that can be tested by low-energy laboratory experiments~\cite{Brax:2017bcp,Brax:2019iut}. Notice that this would also give directly the value
of the cosmological constant at the end of Starobinski's inflation
\begin{equation}
c_{\rm end}\Lambda_{\rm end}=\frac{1}{7}
\end{equation}
as the other contributions to~(\ref{balance}) are negligible. This
determines the product $\Lambda c$ which appears in the
super-potential~(\ref{WW}) leading to Starobinski's inflation. Numerically using (\ref{cinf}) we have $\Lambda_{\rm end}\simeq {8\over 35}\cdot 10^{-10}
\beta^{-2} N^{-2} m_{\rm Pl}^2 $. This determines the energy scale of the cosmological constant during inflation
\begin{equation}
E_{\rm inf}\equiv (\Lambda_{\rm end} m_{\rm Pl}^2)^{1\over 4} \simeq
\left({8\cdot 10^{-10}\over 35\beta^2 N^2}
\right)^{1\over4} m_{\rm Pl}\simeq 9 \times 10^{14} \ {\rm GeV}
\end{equation}
corresponding to a sub-Planckian regime of the effective field theory after compactification  with a cosmological constant $E_{\rm inf}$ close to the Grand Unified scale.

 From a UV point of view, this value only reinforces the fact that the physics at high energy
seems to be largely constrained by the physics at low energy. This is the case of the mass of the scalaron during inflation which is constrained by the CMB data. Here we found
that the physics of the vacuum in the IR, i.e.\ the vacuum stability combined with gravitational tests, determines indirectly the value of the cosmological constant in the UV. Of course, our analysis does not provide any explanation for this value from a top-bottom point of view.

 Finally, let us mention that the narrow interval of mass $3.8\times 10^{-3}\, {\rm eV}\lesssim m_\varphi \lesssim 0.1\times\, {\rm eV}$ is compatible with the
value $m_\varphi\simeq 4.4\times 10^{-3}$ eV for which the scalaron could
be at the origin of the observed dark matter
abundance~\cite{Cembranos:2008gj,Cembranos:2015svp,Shtanov:2021uif,Shtanov:2022xew}. In
this scenario, the coupling of the scalaron to the Higgs field, coming
from the coupling to matter that we have discussed at length, implies
that, at low energy compared to the inflation scale,  the vacuum
expectation value of the scalaron $\varphi$ is displaced from the origin by an amount depending on the electroweak scale $v\sim 250$ GeV\footnote{For more generic initial conditions after inflation taking into account the quantum fluctuations of the scalaron during inflation, the whole interval up to $m_\varphi\simeq 0.1$ eV could accommodate dark matter.}. As the electroweak transition begins, the scalaron starts oscillating with a decreasing amplitude eventually converging to the origin. This misalignment mechanism is similar to what happens for axions and leads an abundance of dark matter which fits the observed value for $m_\varphi \simeq 4.4\times 10^{-3}$ eV~\cite{Shtanov:2022xew}. Combining both scenarios, this would lead to a possible signal in gravitational experiments below a distance $d\lesssim 45\,\mu{\rm m} $. The possibility of testing the existence of a new interaction mediated by the scalaron whose existence could play a role in both dark energy and dark matter is certainly worth pursuing.

\section{Conclusion}\label{sec:conclusion}

The dark sector of the Universe and in particular dark energy could be the result of the gravitational dynamics of the Universe. This could
follow from massive gravity for instance or scalar theories which would mimic the behaviour of the cosmological constant in the late time limit. Another
possibility which has been mostly overlooked is that the dark energy could result from the IR limit of the vacuum energy of a theory whose spectrum would
include at least one light degree of freedom coming from the gravitational sector of the model. This would influence the renormalisation group evolution of the
vacuum energy and could combine its effect to the contributions of the neutrinos to generate the right amount of dark energy. In this article, we consider such
a scenario where the light field is the volume modulus of string
compactifications whose effective field theory at low energy is a
$f({\cal R})$ model of the Ricci scalar squared ${\cal R}^2$  type. This
is true once the volume modulus is stabilised and as long as the curvature of the Universe is lower than the string and compactification scales. Using the reasonable assumption
that the vacuum energy in the IR is never too negative to imply the
collapse of structures in the Universe, we find that the mass of the
scalaron in the IR is tied to the measured cosmological constant in a
way which could be testable with future tests of gravity in the
sub-millimetre range. En route, we describe how the scalaron's,
i.e.\ the volume modulus', effective field theory after
compactification can be described as a $N=1$ supergravity with a
two-parameter family of K\"ahler potentials, including the familiar
no-scale models, and a unique superpotential that we determine its
series expansion  in the large volume limit. The link between the UV
where the scalaron can lead to inflation like in the original
Starobinski model and the vacuum properties in the IR is provided by
the renormalisation group evolution of the scalaron mass
and the vacuum energy (a similar approach has
  been considered in~\cite{Wetterich:1987fm}). In particular,  if the
  scalaron both is responsible from
inflation in the UV, and participates in the dynamics of dark energy in
the form of vacuum energy in the IR, then the scalaron effective field
theory after compactification is almost uniquely determined,
i.e.\ inflation determines the mass of the scalaron and dark energy the
cosmological constant in the UV.  Of course this bottom-up approach
only provides a set of likely constraints on the set of possibilities
for these couplings after compactifications. No dynamical principle
determines their values which are simply fixed by observations.

\acknowledgments
We thank Emilian Dudas and Patrick Valageas for  useful comments.
P.V. thanks the LAPTh for the hospitality during the
completion of this work. P.B. thanks CERN for a research associateship during which parts of this work
were completed.
The work of P.V. has received funding from the ANR grant ``SMAGP''
ANR-20-CE40-0026-01.

\appendix

\section{An explicit example}
\label{ap1}

Our analysis can be exemplified using the effective potential for the volume modulus described in~\cite{Brinkmann:2023eph}
\begin{equation}
V_E({\cal V})= {\cal V}^{-2}\left (U-  \alpha (\ln {\cal V})^{3/2}\right ) + \gamma {\cal V}^{-x}
\end{equation}
in the Einstein frame. Here we have $0<x<2$. In the supergravity frame, this corresponds to the effective potential
\begin{equation}
V_{\rm vac}({\cal V})= {\cal V}_0^{-2}\left (U-  \alpha (\ln {\cal V})^{3\over2}\right )+ \gamma {\cal V}_0^{-2} {\cal V}^{2-x}
\end{equation}
where ${\cal V}_0$ is the minimum of $V_{\rm vac}({\cal V})$ determined by
\begin{equation}
 \frac{3}{2} \alpha (\ln {\cal V}_0)^{1\over2}= (2-x) \gamma {\cal V}_0^{2-x}.
\end{equation}
The effective potential can then be written as
\begin{equation}
V_{\rm vac}({\cal V})= \gamma {\cal V}_0^{-x} \left ( \frac{2(2-x)}{3(\ln {\cal V}_0)^{1\over2}}\left( \frac{U}{\alpha}- (\ln {\cal V})^{3\over2}\right) + \left(\frac{{\cal V}}{{\cal V}_0}\right)^{2-x}\right).
\end{equation}
In the Einstein frame the potential is given by
\begin{equation}
  V_E({\cal V})=   \left(\frac{{\cal V}_0}{\cal  V}\right)^{2} V_{\rm vac}({\cal V}).
\end{equation}
Let us assume that during inflation when the observable scales in the
cosmic microwave background leave the horizon the  volume ${\cal
  V}\approx {\cal V}_\star$  is such that $x \ln \frac{{\cal
    V}_\star}{{\cal V}_0} \ll 1$.
During inflation, the Einstein frame potential reduces to
\begin{equation}
V_{E}(\varphi)\approx  \gamma \left ( 1 -2\left(\frac{\ln{\cal V}_\star}{\ln {\cal V}_0}\right)^{1/2}\frac{{\cal V}_0}{{\cal V}_\star} e^{2\beta\varphi} \right )
\end{equation}
where ${\cal V}/{\cal V}_0= e^{-2\beta \phi}$.
Notice that in that regime the potential in the supergravity frame is simply
\begin{equation}
  V_{\rm vac}({\cal V})  \approx V_{\rm vac}({{\cal V}_*})+
 \gamma  \left ( 2 \left(\ln {\cal
       V}_\star\over \ln {\cal V}_0\right)^{1\over2}- \frac{\ln{\cal V}_\star}{\ln {\cal V}_0}\frac{{\cal V}_0^2}{{\cal V_\star}^{2}}
+ \left(\frac{\cal V}{{\cal V}_0} -\left(\frac{\ln{\cal V}_\star}{\ln {\cal V}_0}\right)^{1\over2}\frac{{\cal V}_0}{{\cal V}_\star}\right)^2\right)
\end{equation}
which is a quadratic potential  with  an  effective minimum that is not situated at ${\cal V}_0$.

We deduce that
\begin{equation}
  \eta=-8\beta^2 \left(\frac{\ln{\cal V}_\star}{\ln {\cal
      V}_0}\right)^{1/2}\frac{{\cal V}_0}{{\cal
      V}_\star}e^{2\beta\varphi_\star}
\end{equation}
and approximately the number of $e$-foldings
\begin{equation}
  N= \frac{1}{8\beta^2} \left(\frac{\ln{\cal V}_0}{\ln {\cal
      V}_\star}\right)^{1/2}\frac{{\cal V}_\star}{{\cal
      V}_0}e^{-2\beta\varphi_\star}
\end{equation}
which determines $n_s-1= -\frac{2}{N}$. Notice that the integral determining $N$ is dominated by the behaviour of the integrand close to $\phi_\star$ where the approximation to the potential is accurate. Consistency implies that
\begin{equation}
\frac{{\cal V_\star}}{{\cal V}_0}= e^{-2\beta \varphi_\star}= 8\beta^2 \left(\frac{\ln{\cal V}_\star}{\ln {\cal V}_0}\right)^{1/2}\frac{{\cal V}_0}{{\cal V}_\star} N
\end{equation}
which determines ${\cal V_\star}/{\cal V}_0$.

Toward the end of inflation ${\cal V}$ differs from ${\cal V}_\star$. On the other hand,   the potential in the supergravity frame  is again quadratic around the true minimum ${\cal V}_0$. The distortion to the quadratic shape affects only  the evolution of the volume modulus between these two epochs of inflation. This will hardly change the relation between the number of $e$-foldings $N$ and the spectral index $n_s$ as $N$  is essentially determined by the shape of the potential around ${\cal V}_\star$. In conclusion,   the potential in the supergravity frame is well approximated by a quadratic form  around ${\cal V}_\star$  during the creation of the observable structures and the inflationary potential by the Starobinski potential.

\section{The renormalisation group}
\label{app:reg}

In the main text, we discuss the evolution of the mass of the scalaron
under the renormalisation group between high and low energies. In
particle physics, the renormalisation scale is usually identified with
the typical energy of a given collision. In the cosmological context
that we have considered, the interpretation of the scale $\mu$ needs
to be discussed more precisely. As we are considering the
renormalisation group in the decoupling subtraction scheme, see for
instance~\cite{Georgi:1993mps} or the recent textbook~\cite[\S7.2.3]{Burgess:2020tbq} for the
methods reviewed here for the decoupling substraction scheme, the scale $\mu$ corresponds to the largest momentum scale for virtual particles running in loops. Particles contribute to the running of the coupling constant as long as they have not been integrated out, i.e. as long as $\mu$ is larger than their mass. When the scale $\mu$ goes through the threshold at the mass $m$, the particle is removed from the particle content of the model whilst a threshold correction is added to the coupling constant. In cosmology, we use this Wilsonian setting in the context of particles such as the ones in the standard model with a typical momentum given by the temperature of the plasma $T$. At each epoch in the history of the Universe, particles are integrated  when the temperature falls below their masses. This allows us to identify $\mu\sim T$.

Let us illustrate the decoupling substraction scheme in the simple case of a massive scalar of mass $m$.
The vacuum energy is given by
\be
\rho= \frac{1}{2} \int \frac{d^3p}{(2\pi)^3} \omega_p.
\ee
where $\omega_p= \sqrt{\vec p^2 +m^2}$.
This  can be written as
\be
\rho= \int \frac{dE d^3p}{(2\pi)^4} \frac{\vec p^2 +m^2}{E^2+\vec p^2 +m^2}.
\ee
It is convenient  to introduce the Euclidean vector $p_E=(E,\vec p)$ and from rotation invariance we have
\be
\int \frac{d^4p_E}{(2\pi)^4} \frac{\vec p^2 }{\vec p_E^2 +m^2}= \frac{3}{4}\int \frac{d^4p_E}{(2\pi)^4} \frac{\vec p_E^2 }{\vec p_E^2 +m^2}
\label{trick}
\ee
Using the 't Hooft-Weltmann regularisation procedure \cite{Zinn-Justin:2002ecy}
$
\int d^4 p_E p_E^n=0, \ n\ne -4,
$
we find that  the vacuum energy is related to the Feynman propagator at coinciding points
\be
\rho= \frac{m^2}{4}G_F(0), \ G_F(0)= \int \frac{d^4p_E}{(2\pi)^4} \frac{1 }{\vec p_E^2 +m^2}.
\ee
We can now calculate
\be
G_F(0)= \frac{S_3}{(2\pi)^4} \int_0^\infty  dp_E\frac{p_E^3}{p_E^2 +m^2}
\ee
where $S_3=2\pi^2$
and finally
\be
G_F(0)= \frac{m^2}{8\pi^2} \int dx \frac{x^3}{x^2+1}.
\ee
Making $x^3= x(x^2+1)-x$ and $\int dx\ x =0$ in dimensional regularisation, we get
\be
G_F(0)= -\frac{m^2}{8\pi^2} \int dx \frac{x}{x^2+1}.
\ee
We now regularise the divergence by applying  a cut-off at a scale $x_{\max}= \frac{\mu^2}{m^2}$ corresponding to a Lorentz invariant cut-off in $p_E \le \mu$, i.e. we only integrate over the quantum fluctuations with momenta up to $\mu$.
We therefore find
\be
\rho(\mu)=-\frac{m^4}{64\pi^2} \ln\left(1+ \frac{\mu^2}{m^2}\right).
\label{exact}
\ee
Two regimes are particularly important
\ba
&& \mu\gg m \ \ \rho(\mu) \simeq  \frac{m^4}{64\pi^2} \ln\frac{m^2}{\mu^2},\cr
&& \mu \ll m \ \ \rho \simeq 0\ .
\label{ren}
\ea
Hence  the particle only participates in the vacuum energy when $\mu\gtrsim m$ as we advocated. This corresponds to the renormalisation group equation
\be 
\frac{d\rho(\mu)}{d\ln \mu}=- \frac{m^4}{32\pi^2} \theta(\mu-m)
\ee
that we have used in the main text. With this we can write the Wilson effective action. When $\mu<m$, the scalar is integrated out and the effective action contains only the vacuum energy
\be 
S_W=- \int d^4x \sqrt{-g} \rho_{\rm vac}
\ee
where $\rho_{\rm vac}$ is the vacuum energy when all the fluctuations have been integrated out, i.e. $\mu\to 0$, and $S_w$ can be identified with the 1PI effective action for vanishing external sources. When $\mu>m$ the Wilsonian action is
\be 
S_W=\int d^4x \sqrt{-g} \left( -\rho_{\rm vac} - \frac{m^4}{64\pi^2} \ln\frac{m^2}{\mu^2} -\frac{(\partial \phi)^2}{2} -\frac{m^2}{2} \phi^2\right).
\ee
This result generalises to the cases in the main text where the renormalisation group allows one to evolve both $\rho(\mu)$ and $c(\mu)$ when  massive particles are integrated out.  This allows us to evaluate the vacuum energy at the end of inflation from the IR vacuum energy and all the contributions from massive particles which are integrated out when the temperature crosses $T=m$. This is the main point used in the paper. 

Although the running of the coupling constant that we used is only between the end of inflation down to low energy, the effective action can also be used during inflation as the physical modes have physical  momenta outside the horizon corresponding to $k\le H_{\rm inf}$ associated to a scale $\mu$  at the Hubble scale. Integrating out the modes inside the horizon with $k>H$ is at the heart of the stochastic description of inflation pioneered by Starobinski \cite{Starobinsky:1994bd}.
Finally for the non-relativistic protons of the ionised hydrogen gas in a  galaxy cluster with momenta $k_p\sim \sqrt{m_p T}$ the scale is then $\mu \sim k_p \sim m_e$ when $T\simeq 1$ keV corresponding to the vacuum energy $\rho(m_e)$ as used in the main text. 

\section{Thermodynamical decoupling of the scalaron}
\label{ap2}

The scalaron could in principle be in thermal equilibrium with the particles in the thermal bath and acquire a large momentum of order $T$. This could happen via the interaction of
 the scalaron  with the thermal bath from the coupling
\begin{equation}
{\cal L}_{2}= - \frac{\beta^2}{2} \frac{m_\Psi}{m^2_{\rm Pl}} \varphi^2\bar \Psi \Psi
\label{inte}
\end{equation}
where $\Psi$ is a massive particle when the Universe has the
temperature $T$. This would lead to chemical equilibrium where
scalarons would be created by annihilation of pairs of fermions.
Considering the radiation era where $\Psi$ is relativistic, the cross-section for $\varphi+\varphi \to \Psi+\bar\Psi$
\begin{equation}
  \sigma\sim\left|\begin{gathered}    \begin{tikzpicture}[baseline=(x)]
        \begin{feynman}
       \vertex (x);
       \vertex[above left=1cm of x](c1lu);
       \vertex[below left=1cm of x](c1lb);
       \vertex[above right=1cm of x](c1ru);
       \vertex[below right=1cm of x](c1rb);
      \tikzfeynmanset{every vertex=dot}
            \diagram* {
             (c1lu) -- [scalar, very thick] (x);
        (c1lb) -- [scalar, very thick] (x);
        (x) -- [fermion, very thick] (c1rb);
          (c1ru) -- [fermion, very thick] (x);
                                };
                              \end{feynman}
                    \end{tikzpicture} \end{gathered}\right|^2 ,
\end{equation}
is of order $\sigma\simeq \beta^4 \frac{m^2_\Psi}{m^4_{\rm Pl}}$ and the interaction rate which could maintain thermal equilibrium is, for the relativistic particles $\Psi$, given by $\Gamma \sim g T^3 \sigma$ where $g$ is the number of relativistic species. The chemical equilibrium is maintained as long as the reaction rate is larger than the Hubble rate and no cosmological dilution takes place, i.e.\  $\Gamma >H\sim \frac{\sqrt{g} T^2}{m_{\rm Pl}}$ corresponding to the bound
\begin{equation}
T \gtrsim T_{\rm dec}= \frac{1}{\sqrt g \beta^4}\frac{m^3_{\rm Pl}}{m^2_\Psi}
\end{equation}
with a decoupling temperature typically larger than the Planck scale. As a result,   the scalaron is never in thermal equilibrium with the thermal bath.

Elastic processes could also raise the typical momentum of scalarons to a value of order $T$. This follows from the Yukawa coupling
\begin{equation}
{\cal L}_{\rm yuk}= -\frac{ \beta m_\Psi}{m_{\rm Pl}} \varphi \bar \Psi\Psi
\label{yuk}
\end{equation}
allowing for the  kinetic reaction $\varphi+ \Psi \to \varphi +\Psi$
mediated by $\Psi$.
\begin{equation}
 \sigma\sim\left|\begin{gathered}     \begin{tikzpicture}[baseline=(x)]
        \begin{feynman}
       \vertex (x);
       \vertex[above left=2cm of x](c1lu);
       \vertex[below left=2cm of x](c1lb);
       \vertex[above right=2cm of x](c1ru);
       \vertex[below right=2cm of x](c1rb);
      \tikzfeynmanset{every vertex=dot}
        \vertex [left=1cm  of x] (xl);
        \vertex [right=1cm  of x] (xr);
            \diagram* {
             (c1lu) -- [scalar, very thick] (xl);
        (c1lb) -- [fermion, very thick] (xl);
        (xr) -- [fermion, very thick] (c1rb);
          (xr) -- [scalar, very thick] (c1ru);
         (xl)  -- [fermion, very thick] (xr);
                                };
                              \end{feynman}
                    \end{tikzpicture}\end{gathered}\right |^2.
\end{equation}

This could take scalarons with initially low momenta such as $k_{\rm ex}(T)$ and by  momentum transfer of order $T$ leads to momenta for the scalarons of order $k\sim T$.  The interaction rate $\Gamma= g  \beta^4 \frac{m^4_\Psi}{m^4_{\rm Pl}} T $ is very small implying that there is far less than one interaction per Hubble time. As a result, the scalarons never receive a momentum transfer of order $T$.

\end{document}